\documentclass[12pt]{article}
\input{epsf.tex}
\begin{document}
\begin{center}
{\bf ON THE CORRECTNESS OF VARIOUS APPROACHES IN THE
EXTRACTION OF THE NUCLEUS PARAMETERS ON EXAMPLE OF ANALYSIS OF THR TWO-STEP
$\gamma$-CASCADES IN $^{163}$Dy COMPOUND NUCLEUS}

{\bf V.A. Khitrov, A.M. Sukhovoj}\\
{\it Frank Laboratory of Neutron Physics, Joint Institute
for Nuclear Physics, 141980, Dubna, Russia}\\

{\bf Pham Dinh Khang}\\ 
{\it National University of Hanoi}\\
{\bf Vuong Huu Tan,  Nguyen Xuan Hai}\\
{\it Vietnam Atomic Energy Commission}\\ \end{center}

\section{Introduction}

All results of the analysis of experiment must be reliable,
i.e., even having inevitable systematic errors, they must not lead
to deliberately false conclusions about the nature of the studied phenomenon.
For example, any selection and subsequent parameterization of nuclear models must
provide the minimal probability of the appearance of wrong result,
and must not lead to the generation of false hypotheses.

The greatest probability of obtaining the deliberately erroneous ideas about
the studied process occurs at the stage of the extraction of its parameters
from the measured data. First of all this is manifested in case of indirect
experiments. 
The obvious example of this possibility is the extraction of information on
the parameters of the compound-state cascade gamma-decay for nucleus with the
high level density from the analysis of the two-step cascade intensities or
any other measured spectra.

It is characteristic for these cases, that the dependence the intensity of the measured
distribution on the level density $\rho$ and the radiative strength
functions $k$ of the emission of nuclear reaction products is nonlinear.
The corresponding systems of equations either are degenerated, or are close to
the degenerated one's. 
And the existing ideas about the unknown parameters to greater or lesser
extent are approximate or do not take into account the whole nuclear
information accumulated by to now. The above mentioned ideas are well illustrated
by the analysis of experimental data of the two-step cascade
intensities occurring at  thermal neutron capture in $^{162}$Dy.
Experience of study of  two-step cascade intensities study by the Dubna group in
more than 50 nuclei in the region $28 \leq A \leq 200$
(approximately a fourth of these data were obtained by V. Bondarenko group
in Riga and J. Honz\'atko  group in R\'ezh) allows us to assert
that their analysis can either ensure obtaining or maximum reliable
(at present) conclusions on the process of the cascade
gamma-decay of nucleus with any highest density of levels [1,2],
or lead to the completely erroneous conclusions on this studding process.

\section{Specific character of obtaining experimental data for $\rho$ and $k$
in the nuclei with the high level densities}

Experimental intensities of the two-step cascades between the neutron resonance
and the group of low-lying levels from $E_f < 1$ MeV region are measured
by Ge-detectors, i.e., the sequence of quanta in these cascades
cannot be determined by apparatus. Accordingly, in any interval $\Delta E$
of cascade gamma-quanta energy $E_\gamma$ the experimental intensity 
$I_{\gamma\gamma}^{exp}=I_{\gamma\gamma}(E_1)+I_{\gamma\gamma}(E_2)$
is the sum of unknown terms,  corresponding either to the primary,
or to the secondary cascade transition. 
Their energies $E_1$ or $E_2$ in this case satisfy the obvious conditions:
$E_\gamma \leq E_1 \leq E_\gamma+\Delta E$ and
$E_\gamma \leq E_2 \leq E_\gamma+\Delta E$.
Taking into account the physical limitation,
it is only possible to conclude, that the cascade intensity with the fixed order
of the cascade quanta always lies in the interval of its value from zero to
$I_{\gamma\gamma}^{exp}$.
And this circumstance leads to a catastrophic increase [3] of the interval widths
both values $\rho$ and $k$, which can reproduce $I_{\gamma \gamma}^{exp}$ for
all possible values of energies of the cascade gamma-transitions
$E_1+E_2=B_n-E_f$ with one and the same $\chi^2$.
Moreover - not only for the cascades for one fixed low-lying level,
but also for their set for several final levels together.

This problem can be solved only by determining the cascade intensities
in the function of their primary gamma-transition energy $E_1$:
\begin{equation}
 I_{\gamma\gamma}(E_1)=\sum_{\lambda ,f}\sum_{i}\frac{\Gamma_{\lambda i}}
{\Gamma_{\lambda}}\frac{\Gamma_{if}}{\Gamma_i}=\sum_{\lambda ,f} 
\frac{\Gamma_{\lambda
i}}{<\Gamma_{\lambda i}> m_{\lambda i}} n_{\lambda 
i}\frac{\Gamma_{if}}{<\Gamma_{if}> m_{if}}. 
\end{equation} 
The functional (1) depends both on the ratios of partial and total radiative widths
$\Gamma i$ of primary $E_1$ and secondary $E_2$ gamma-transitions in the
cascades between the levels $\lambda$, $i$ and $f$, and on the number
$n(m)=\rho \times \Delta E$
of levels excited in this case in different energy intervals.

The practical possibility of determining this intensity was proposed by the
authors [4]. The most probable error in the operation of division the experimental
spectrum into the parts $I_{\gamma \gamma}(E_1)$ and $I_{\gamma\gamma}(E_2)$
with good precision can be determined from the approximation [5] in the
assumption, that the distributions of random values of the cascade intensities
with the dipole electric and magnet primary cascade transitions
are described by a pair of distributions with some averages and by dispersion
slightly exceeding the dispersion of the Porter-Thomas distribution.

The degree of detailing, with which can be
extracted  from observed $I_{\gamma \gamma}(E_1)$
the form of energy dependence of both level density, and radiative
strength functions is determined by the optimum width of averaging interval $\Delta E$
of the cascade intensities.
A good compromise between the real possibilities of contemporary detector
and computer technology and the inevitable fluctuations of partial widths is
reached at the width of interval $\Delta E=$ 50 - 100 keV or larger, into which the
excitation energy of nucleus is divided .
Type of successive transition (dipole electrical or magnetic one), spin and parity
of the intermediate level $i$ excited in this case are uniquely determined by
known values $J^{\pi}$ levels of $\lambda$ and $f$.
A very small intensity of the cascades between the levels with spin difference
$|J_\lambda - J_f |>2$ exclude the need for to use the
transitions of the highest multipolarities in the analysis [1,2].
A possibility of their including in the analysis does not cause doubts, howerer,
practical expediency is completely absent. 

In the cascades to the levels with the spins $|J_\lambda - J_f | \leq 2$
the admixture of E2-transitions is taken into account [1,2] automatically due to the
appropriate distortion of the radiative  strength functions of pure
M1-transitions.
The direct calculation of gamma-transitions of the highest multipolarities
is required in this case only to reproduce the intensity of cascades
to the level $E_f=73$ keV with the spin $J=7/2$, the intensity of which is less
than errors of the remaining values of $I_{\gamma\gamma}$.
And the including of the additional free parameter in the analysis with
conservation of the number of the experimental points will unavoidably increase
the uncertainty of the obtained values $\rho$ and $k$.

For convenience of the direct comparison the obtained radiative
strength functions and E1-, and M1-transitions in the nuclei of
different mass $A$ it is expedient to determine in the form:
\begin{equation} 
k=\Gamma_{\lambda i}/(E_{\gamma}^3\times
A^{2/3}\times D_{\lambda})
\end{equation}
and to use
the experimentally measured and approximated into [6] near $B_n$ ratios
$k(M1)/k(E1)$ for mutual standardization in the iterative processes [1,2].

Calculation of the internal conversion of gamma-quanta and form
of Ge-detector line at the registration threshold $E_\gamma > 520$ keV
is not required.  The intensity of cascades in the expression (1) is proportional
to the derivative $dk/>dE$ and, in the first approximation, is inversely
proportional to $\rho$.  This ensures the maximum sensitivity of the experiment in
the region of the smallest values of $\rho$.
I.e., in the excitation energy regions, where the influence of nuclear
structure on the parameters of the investigated nuclear reaction must be
a maximum. A fundamentally different form of the relation between the unknown
parameters in comparison with the usual evaporative ones and the gamma-spectra
provides the smallest influence of correlation of the parameters on
their real error.

Interval of the intermediate level density values  and partial radiative
widths of cascade gamma-transitions, reproducing an experiment with the assigned
accuracy,  can be determined from (1) with the smallest possible systematic
error only with using [2] the additional experimental information about the
relation of the partial widths of the primary and secondary transitions of one
and the same multipolarity and energy;
values of the total radiative width of compound-state [7] and the information
about the density of low- [8] and high-lying [7] levels
(here - neutron resonances).

The values $\rho$ and $k$ obtained from the system of equations (1) are very rigidly
determined (interval of their possible values is always small) and have
principally unremoval deviations with the results obtained within the
framework of other known procedures [9,10] and the conventional ideas
[11-14] on this nuclear parameters.

\section{Specific character of the existing model notions for the nuclei
with high level density}

Actually, the only basis of the existing model notations about $\rho$ used in  practice
by experimenters is the notion of  nucleus as a system of noninteracting Fermi gas [11]
and the conclusion following from it
about a smooth change in the level density with an increase
in the excitation energy of nucleus.
Like the Gilbert-Cameron model [12] analogous in essence,
it completely ignores the existence of the excitations of
vibration (and rotation - for the deformed ones) types in the nucleus and their interaction
with quasi-particles. Therefore the indicated model notions cannot
but give a significant error in the predicted values of $\rho$.

Unfortunately, this situation was not  changed much by present-day
models of the level density. For example, the general model of superfluid nucleus
[13] directly takes into account the mutual transformation of nucleus from the
superfluid state into the usual ones.
However, its parameters are selected (and fixed into the model) on the basis
of experimental information about $\rho$, containing unknown but significant
systematic errors. First of all, it may say on the data about $\rho$, which obtained from the
evaporated nucleons spectra.  Usually authors of the corresponding
experiments carry out a subjective selection of the potentials used in the
optical nuclear model to calculate the probability of emission
of the reaction product by an excited nucleus.
Corresponding experimental data are quite absent up to now and resulted systematic
error cannot be determine even in principle.

Criteria of the corresponding selection are not published,
therefore it is only possible to assume, that it is just a degree of ``smoothness"
of the obtained dependence $\rho=f(E_{ex})$.
Similar selection of versions of analysis is also inevitable in the procedure
of extraction of $\rho$ and $k$ from the spectra of primary gamma-transitions
during decaying of the levels with different energy of their excitation in the
reactions of the type of $(d,p)$ or $(^3$He,$\alpha,\gamma)$.
Its need is caused both  by the enormously high coefficients of the spectrum 
errors transfer to the errors of the determined parameters 
$\rho$ and $k$, and by other specific sources of significant systematic errors. 

Parametrization [16] of model of the partial level density for $n$-quasiparticles
[17] showed, that the contemporary model
notions about the nucleus are quite capable to reproduce ``non-smooth" functions
$\rho=f(E_{ex})$ with high accuracy. It is also not contradict to the
known physical picture of interactions and the mutual transformation of boson
and fermionic macrosystems into each other.

\section{The two-step cascade experimental data analysis in $^{163}$Dy}

Experimental data on the intensities of the two-step cascades from the
$^{162}$Dy$(n,\gamma)$ reaction [18] have been obtained by us with substantially
smaller statistics, than in [19] and with a slightly  different
positions of sample and Ge-crystals. 
These circumstances can distort in different ways,
first of all, the value of the absolute intensity of cascades.
Comparison of the absolute values of their sum from the interval
$(E_1+E_2)/2 \pm 1$ MeV (\% to the decay) for the data [18,19] is presented
in the table. The maximum value of $E_f$ in it corresponds to the data [18], and is
2 times less than in [19] because of the substantial difference of the
efficiencies of the detectors used and intensity of thermal neutron beams.

Table.
\begin{center}
\begin{tabular}{|r|r|r|r|} \hline 
$E_1+E_2$, keV &$E_f$, keV& $I_{\gamma\gamma}$ [18] & $I_{\gamma\gamma}$
[19]\\\hline
6272 & 0   & 3.7 & 3.3\\
6198 & 73  & 0.4 & 0.2\\
6020 & 251 & 2.1 & 2.2\\
5920 & 351 & 3.8 & 3.3\\
5881 & 389 & 3.1 & 2.9\\
5847 & 421+427 & 3.9 & absent\\\hline
\end{tabular}
\end{center}
As seen from these data, the observed difference between the
experimental data cannot substantially [20] influence on the values $\rho$
and $k$, which reproduce precisely [21] the intensities of cascades.
Consequently, the fundamental incompatibility of conclusions obtained from
the different variants [18,19] of analysis indicates the incorrectness, at least,
of one of them. 

This situation is clearly illustrated in Figs. 1-3 by results of the
search for values of $\rho$ and $k$,  maximally accurately reproducing experimental
intensities of two-step cascades in 5 spectra measured by us
(cascades of $E_1+E_2=6198$ keV are not included in the analysis in view of
circumstances mentioned above).

It follows from these data, that the experimental spectra of $^{163}$Dy with
the unknown sequence of quanta in the cascades can be reproduced with the
maximally high accuracy by an infinite number of different functional dependencies
of $\rho$ and $k$.
The ratio of the maximal and of the minimal possible values of these parameters
reaches $\sim 100$. 
The reason for this spread is obvious: for example, very great significances
of $k$ in the region of energy $E_1 \sim 1$ MeV appear because
the corresponding experimental intensities are approximated in a few cases
as corresponding only (or in general only) to primary gamma-transitions.
Naturally, that to define one of the terms from the known value of
their sum the additional experimental data is necessary!)
And with such enormous spread of the values of the parameters one should take
into consideration, that the interval of possible values of $\rho$ and $k$
in Figs. 2,3 (because of the small number of realized random processes [1])
is determined very approximately.
But it cannot be less, than that follows from the data on Figs. 2,3.

Approximation of the of $\rho$ obtained in [21] (with the aid of the procedure [1]
from the values $I_{\gamma\gamma}(E_1)$ only) 
by partial densities of
1-, 3-, 5-, 7- quasi-particle excitations [16]
is represented in Fig. 4 for the version of the logarithmic dependence
of correlation functions on the
excitation energy of nucleus.
The comparison of the sums of radiative  strength functions with the relative
percentages of $n$-quasiparticle levels is executed in Fig. 5.
The coefficient of a vibration increase of the level density in the
case of $^{163}$Dy is very close to other nuclei data and equal to 13.
This means, that more than 90\% of levels lower than the threshold of the
appearance  of 5-quasi-particle excitations has wave functions 
with significant components of vibration type.
However the concrete type of phonons, cannot be determined only from the
intensities of cascades without the
attraction of additional information.

\section{Conclusion}
The model-free analysis of experimental data on the intensities of two-step
cascades in $^{163}$Dy shows, that the basic special features of the cascade
gamma-decay of its compound-state completely correspond to those observed in
other nuclei. And they do not require the introduction of some additional
hypotheses.  Moreover the reliable picture of the studied process can be
obtained only from the distributions of the intensities of cascades for the
determined values of their primary gamma-transition energies .

\begin{flushleft}
\newpage
{\large\bf References}\end{flushleft}\begin{flushleft}
\begin{tabular}{r@{ }p{5.65in}} 
$ [1]$ & E.V.  Vasilieva, A.M.  Sukhovoj, V.A.  Khitrov, Phys.  At. Nucl. 
64(2) (2001) 153, nucl-ex/0110017\\
$ [2]$ &   A.M.  Sukhovoj, V.A.  Khitrov, Par. and Nucl.,  36(4) (2005) 697.\\
$ [3]$ & A.M.  Sukhovoj, V.A.  Khitrov, Li Chol, in Proceedings of
XII International Seminar on   Interaction of Neutrons with Nuclei, E3-2004-
169, Dubna, 2004, p.438, nucl-ex/0409016\\
$ [4]$ &   S.T. Boneva, V.A.  Khitrov, A.M.  Sukhovoj, Nucl.  Phys.  A589 (1995) 293.\\
$ [5]$ &A.M.  Sukhovoj, V.A.  Khitrov, Physics of Atomic Nuclei 62(1) (1999) 19.\\
$ [6]$ & J. Kopecky,
 Proc. of the Symposium on Neutron Capture
Gamma-Ray Spectroscopy
and Related Topics 1981. Ed by von Egidy T. Bristol-London, 1981. P.423.\\
$ [7]$ & S.F.  Mughabghab, Neutron Cross Sections
 BNL-325.  V.  1.  Parts A, B, edited by Mughabhab S.  F., Divideenam M., Holden 
N.E., N.Y. Academic Press, (1984)\\
$ [8]$  & http://www.nndc.bnl.gov/nndc/ensdf.\\
$ [9]$ & O.T.  Grudzevich et.al., Sov.  J.  Nucl.  Phys. 53 (1991) 92.\\
& B.V. Zhuravlev, Bull. Rus. Acad. Sci. Phys. 63 (1999) 123.\\ 
$ [10]$ & G.A.  Bartholomew et al., Advances in nuclear physics 7 (1973) 229.\\ 
& A. Schiller et al., Nucl.  Instrum. Methods Phys.  Res. A447 (2000) 498.\\
$ [11]$ &W. Dilg, W.  Schantl, H.  Vonach, M.  Uhl, Nucl.  Phys. A217 (1973) 269.\\
$ [12]$ &A. Gilbert, A.G.W. Cameron, Can. J. Phys. 43 (1965) 1446.\\
$ [13]$ &S.G. Kadmenskij, V.P. Markushev, V.I. Furman, Sov. J. Nucl. Phys. 37 (1983) 165.\\
$ [14]$ & P. Axel,  Phys. Rev. 1962. 126. $N^o$ 2. P. 671.\\
$ [15]$ &E.M. Rastopchin, M.I. Svirin, G.N.  Smirenkin, Yad. Fiz. 52 (1990)1258.\\
$ [16]$ & A.M.  Sukhovoj, V.A.  Khitrov, JINR preprint E3-2005-196\\
$ [17]$ & V.M.  Strutinsky,  in Proc. of  Int.  Conf.  Nucl. Phys., Paris (1958) 617.\\
$ [18]$ &S.T. Boneva et al., Izv. AN SSSR, Ser. Fiz., (1986) {\bf 50} 1832\\
$ [19]$ & F.Becvar et al., Phys.Rev., (1995) {\bf C52} 1278\\
&F. Becvar et al.,
Eleventh International Symposium on Capture Gamma-Ray Spectroscopy
and Related Topics, Pruhonice, September 2-6, 2002, World Scientific,
Ed. J.Kvasil, P. Cejnar, M.Krticka, p. 726\\
$ [20]$ & V.A.  Bondarenko et all, in Proceedings of XII
 International Seminar on Interaction of Neutrons with Nuclei,
 E3-2004-169, Dubna, 2004, p.  38, nucl-ex/0406030\\
$ [21]$ & V.A. Khitrov,  Li Chol, A.M. Sukhovoj,
In:  XI International Seminar on Interaction
of Neutrons with Nuclei,  Dubna, 22-25 May 2003,
E3-2004-9, Dubna, 2004, p. 92.\\
\end{tabular} 
\end{flushleft}

\begin{figure} [htbp]
\begin{center}
\leavevmode
\epsfxsize=14.5cm
\epsfbox{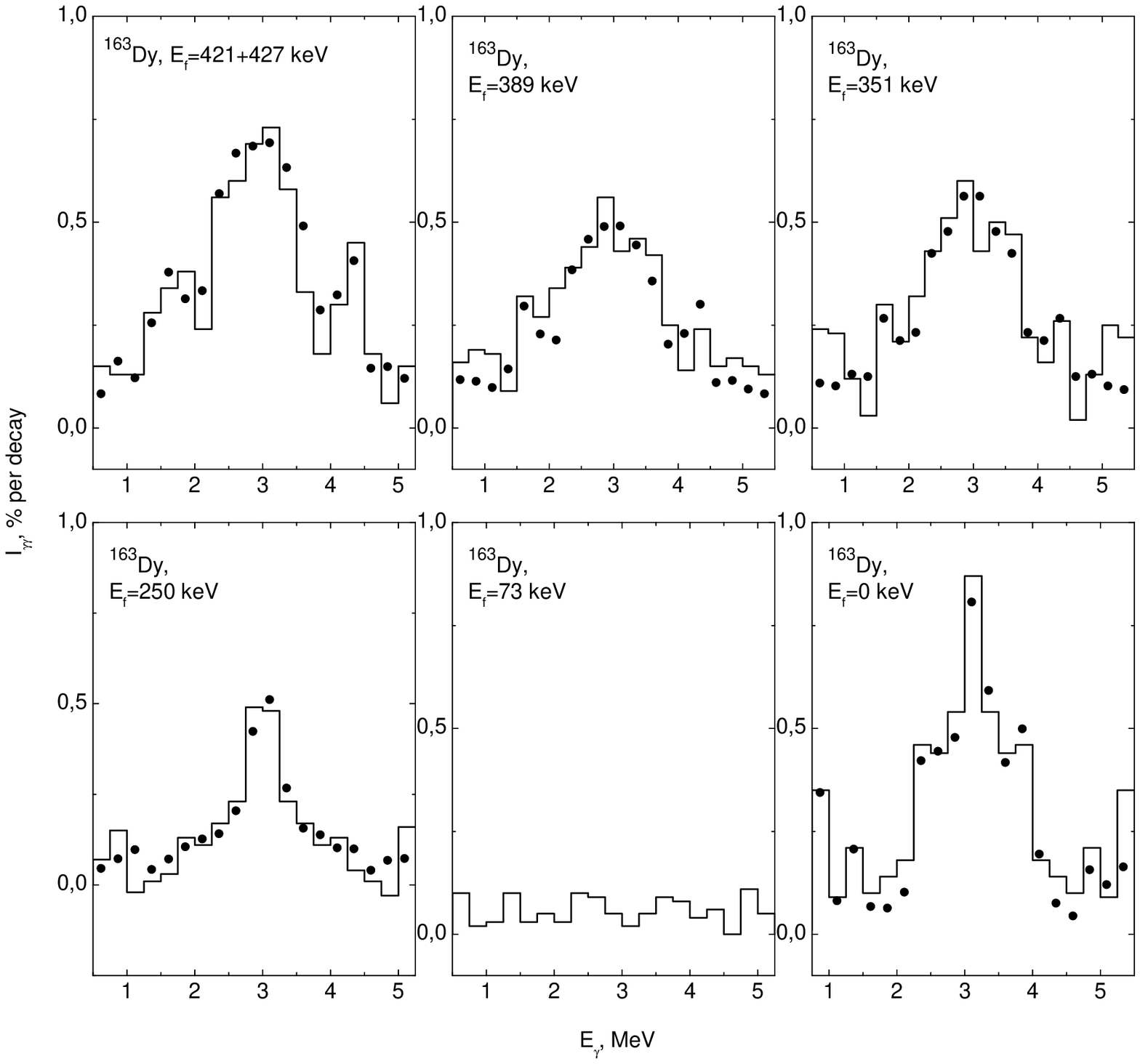}
\end{center}
\hspace{-0.8cm}\vspace{-6.cm}

{\bf Fig.~1} Lines are the total experimental intensities (in \% per decay) of 
two-step cascades (summed in energy bins of 250 keV) as a function of
any cascade quanta energy (is multiplied by 2).
Points - typical approximation for any pair of the level density and radiative
strength functions from those given in Figs. 2,3.
\end{figure}

\begin{figure}
\vspace{1 cm}
\begin{center}
\leavevmode
\epsfxsize=12.5cm
\epsfbox{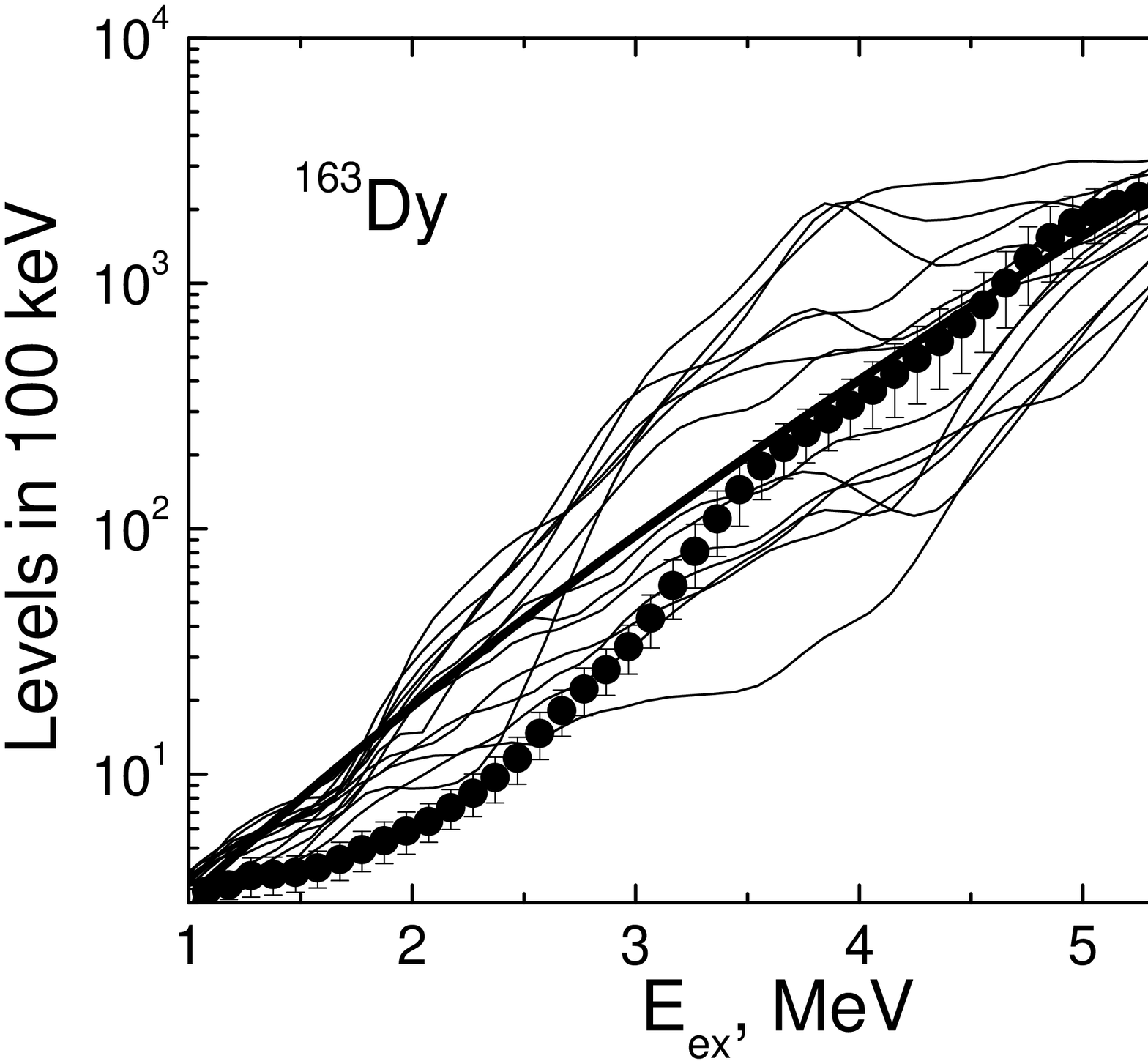}
\end{center}
\hspace{-0.8cm}\vspace{-4.5 cm}

{\bf Fig.~2}~Points with error bars are the interval of probable values of
the level density enabling
the reproduction of the experimental cascade intensity (as a function of
primary transition energy) and total radiative 
width of capture state. The thick line represents predictions of the model [11].
Thin lines - versions of the obtained random functions reproducing
the intensities from Fig. 1.
\end{figure}

\begin{figure}
\begin{center}
\leavevmode
\epsfxsize=12.5cm
\epsfbox{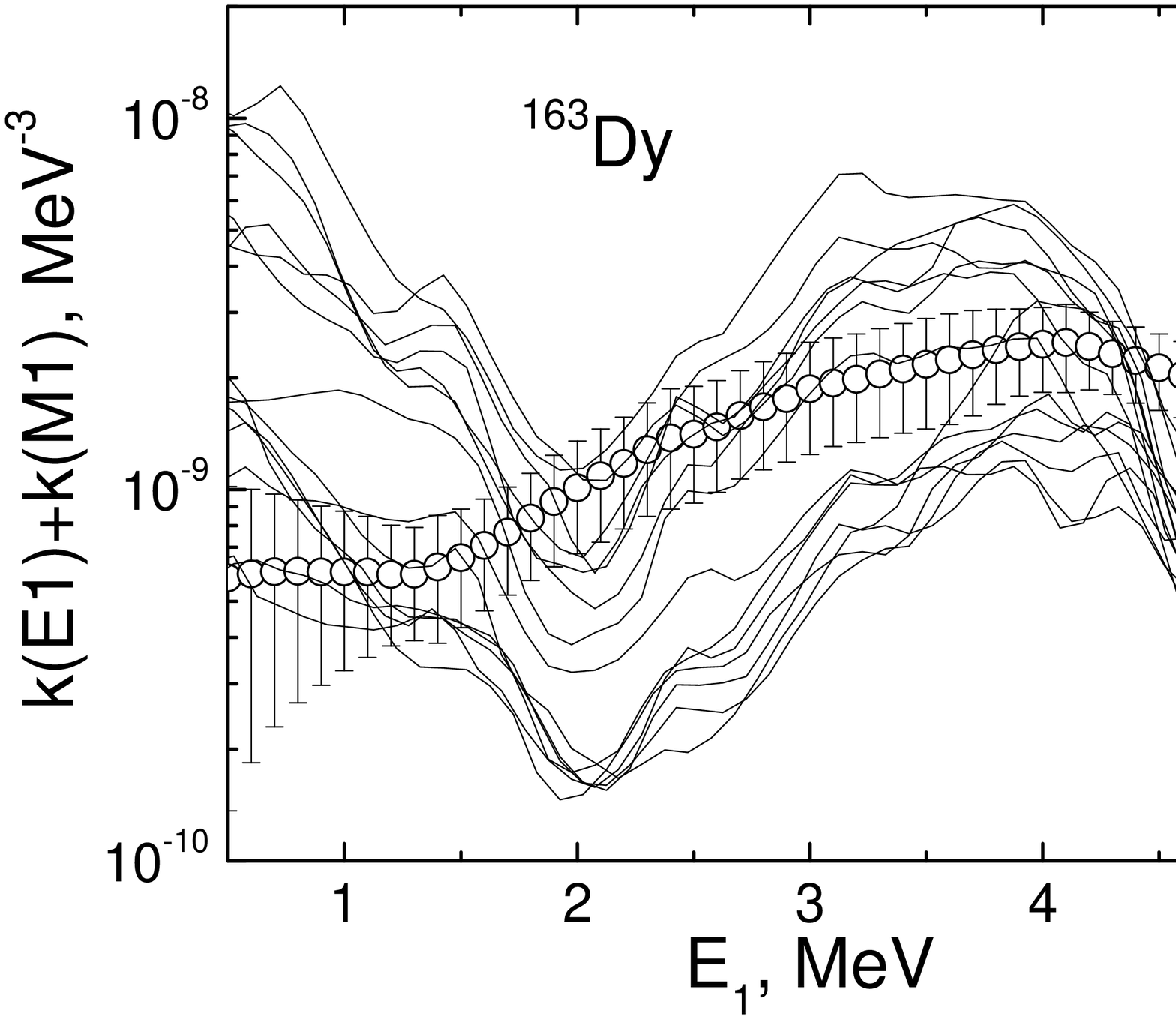}
\end{center}
\hspace{-0.8cm}\vspace{-4.5cm}

{\bf Fig.~3} The probable interval of the sum radiative strength function $k(E1)+k(M1)$
(points with error bars) providing the reproduction of the experimental data.
Thin lines - versions of the obtained random functions reproducing
the intensities from Fig. 1.
\end{figure}
\newpage
\begin{figure}
\vspace{1 cm}
\begin{center}
\leavevmode
\epsfxsize=12.5cm
\epsfbox{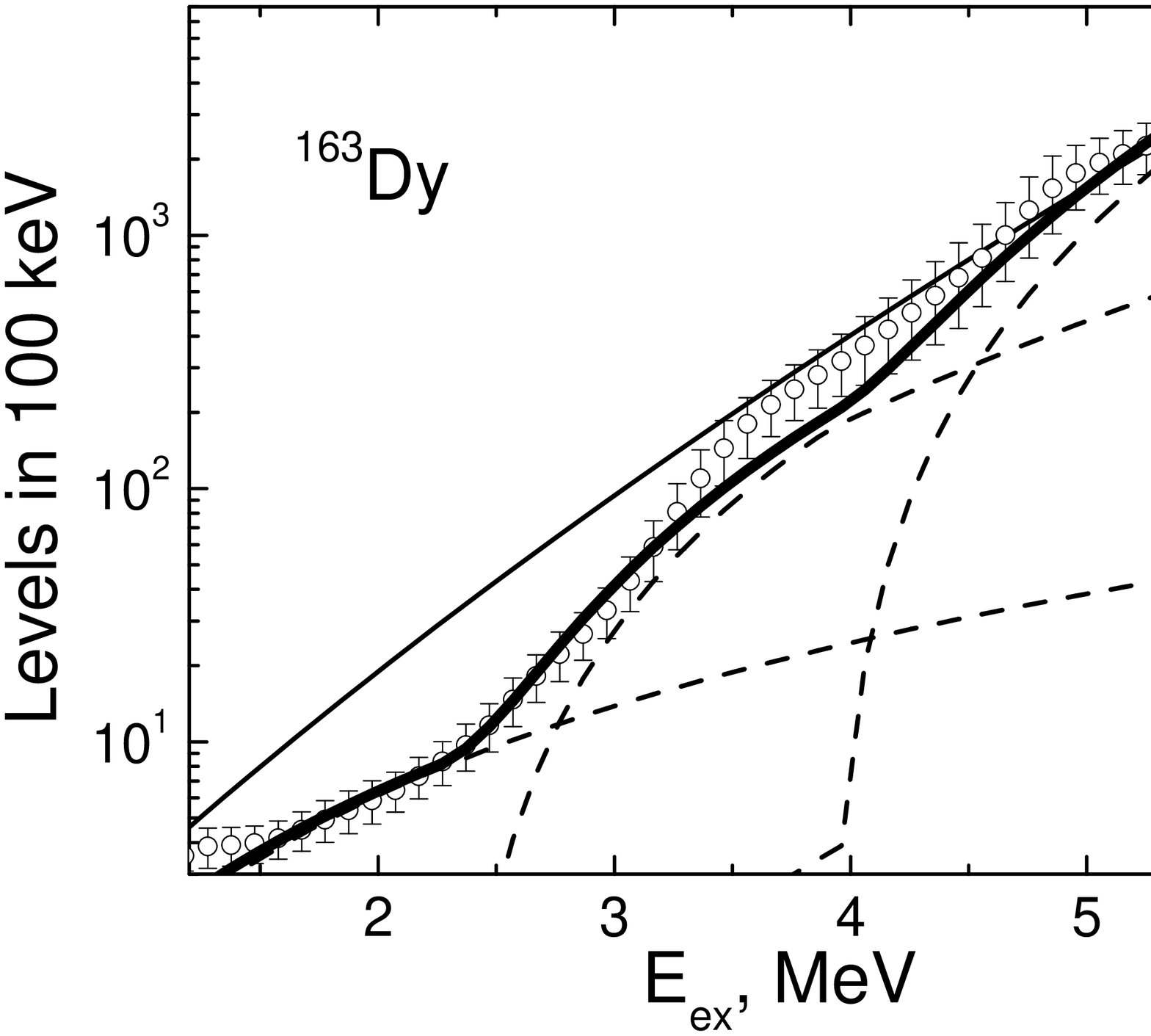}
\end{center}
\hspace{-0.8cm}\vspace{-4.5cm}

{\bf Fig.~4} Dotted line - approximation of the most probable $^{163}$Dy
level density by the partial densities of 3-,  5- and 7- quasi-particle
excitations with the most probable coefficient of its collective 
enhancement factor.  Thick line - sum of partial densities. Thin line is
prediction of the model [11].
\end{figure}

\begin{figure}
\begin{center}
\leavevmode
\epsfxsize=12.5cm
\epsfbox{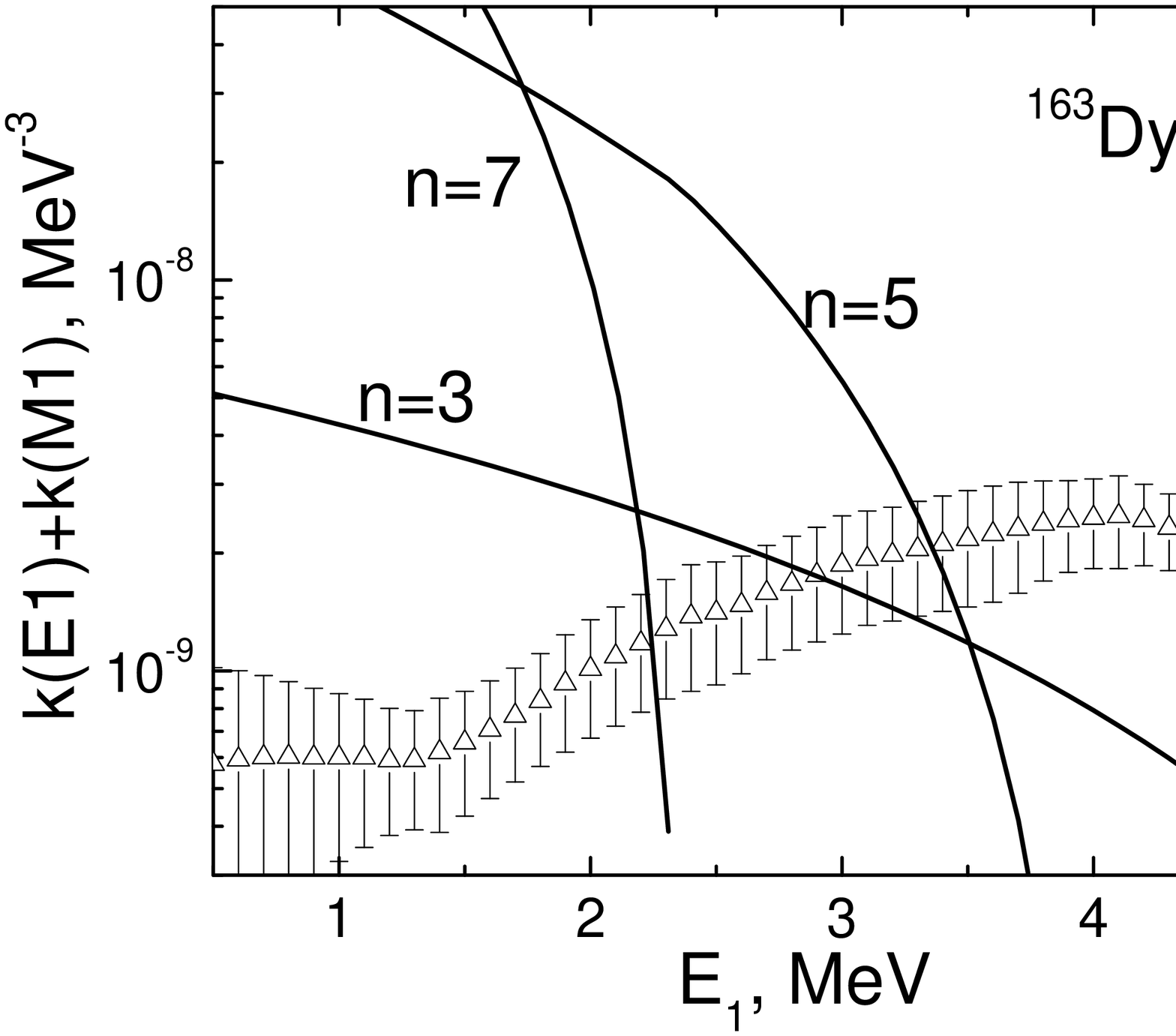}
\end{center}
\hspace{-0.8cm}\vspace{-4.5cm}

{\bf Fig.~5} Points with the errors - most probable sum of radiative
strength functions. Dotted line - relative percentage of levels, corresponding
to excitation $n$-quasiparticles for the energy of their excitation
$E_i=B_n-E_1$.
\end{figure}
\end{document}